\documentclass[twocolumn,aps,showpacs,prl,superscriptaddress]{revtex4-1}
\usepackage{amsmath}
\usepackage{amssymb}
\usepackage{esint}
\usepackage{graphicx}
\usepackage{times}
\usepackage{etoolbox} 

\usepackage{pdfpages}
\makeatletter
\patchcmd{\@outputpage@head}{\@ifx{\LS@rot\@undefined}{}{\LS@rot}}{}{}{}
\makeatother

\usepackage[normalem]{ulem}
\usepackage[colorlinks=true,allcolors=blue,bookmarks=false,pdfusetitle]{hyperref}
\usepackage{url}

\usepackage{color}

\makeatother
\begin{document}
\flushbottom
\title{What is a quantum shock wave?}
\author{S. A. Simmons}
\address{School of Mathematics and Physics, University of Queensland, Brisbane, Queensland 4072, Australia}
\author{F. A. Bayocboc, Jr.}
\address{School of Mathematics and Physics, University of Queensland, Brisbane, Queensland 4072, Australia}
\author{J. C. Pillay}
\address{School of Mathematics and Physics, University of Queensland, Brisbane, Queensland 4072, Australia}
\author{D. Colas}
\address{School of Mathematics and Physics, University of Queensland, Brisbane, Queensland 4072, Australia}
\address{ARC Centre of Excellence in Future Low-Energy Electronics Technologies, University of Queensland, Brisbane, Queensland 4072, Australia}
\author{I. P. McCulloch}
\address{School of Mathematics and Physics, University of Queensland, Brisbane, Queensland 4072, Australia}
\author{K. V. Kheruntsyan}
\address{School of Mathematics and Physics, University of Queensland, Brisbane, Queensland 4072, Australia}

\date{\today}
\begin{abstract}
Shock waves are examples of the far-from-equilibrium behaviour of matter; they are ubiquitous in nature, yet the underlying microscopic mechanisms behind their formation are not well understood. Here, we study the dynamics of dispersive quantum shock waves in a one-dimensional Bose gas, and show that the oscillatory train forming from a local density bump expanding into a uniform background is a result of quantum mechanical self-interference. The amplitude of oscillations, i.e., the interference contrast, decreases with the increase of both the temperature of the gas and the interaction strength due to the reduced phase coherence length. Furthermore, we show that vacuum and thermal fluctuations can significantly wash out the interference contrast, seen in the mean-field approaches, due to shot-to-shot fluctuations in the position of interference fringes around the mean. 
\end{abstract}
\maketitle

\emph{Introduction.}---The study of dispersive shock waves in superfluids, such as dilute gas Bose-Einstein condensates (BECs), has been attracting a growing attention in recent years (see, \emph{e.g.}, Refs. \cite{Hoefger2016_review,Hau2001,Damski2004,Simula2005,Cornell2006,Engels_PRL_2008,Davis2009,Abanov_PRA_2012,Bulgac2012,Peotta2014}). 
This is partly due to the fact that shock waves represent examples of far-from-equilibrium phenomena, for which a fundamental understanding of the laws of emergence from the underlying many-body interactions is generally lacking. Ultracold atomic gases offer a promising platform for addressing this open question due to the high level of experimental control over the system parameters and the dynamics. Other physical systems in which dispersive shock waves form, and which may benefit from such an understanding, include rarefield plasma \cite{Taylor1970_Plasma,Tran1977_Plasma}, intense electron beams \cite{Mo2013_electron_beams}, liquid helium \cite{Rolley2007_liquid_helium}, and exciton polaritons \cite{dominici_polariton_2015}. 

Dispersive (or non-dissipative) shock waves in fluid dynamics are identified by density ripples or oscillatory wave trains whose front propagates faster than the local speed of sound in the medium; 
a typical scenario for their formation is the expansion of a local density bump into a nonzero background. Dissipative shock waves, on the other hand, are characterised by a smooth, but steep (nearly discontinuous) change in the density \cite{Hoefger2016_review,Dissipative_2013,mossman2018dissipative}. 
In either case, the effects of dispersion or dissipation prevent the unphysical hydrodynamic gradient catastrophe 
by means of energy transfer from large to small lengthscales, or through the release of the excess energy via damping.

While dissipative shock waves involve irreversible processes that can be well described within classical 
hydrodynamics, dispersive shock waves in BECs require quantum or superfluid hydrodynamics for their description. The latter can be derived from the mean-field description of weakly interacting BECs via the Gross-Pitaevskii equation (GPE) \cite{dgps99,ps14}. The effect of dispersion is represented here by the so-called quantum pressure term, hence the use of an alternative term for dispersive shock waves---quantum shock waves \cite{Hau2001}. We point out, however, that essentially the same phenomenon can be observed in classical nonlinear optics \cite{Shock_Optics_1,wan2007dispersive_optics}, wherein the electromagnetic dispersive shock waves are generated in a medium with a Kerr-like nonlinearity and are described by the nonlinear Schr\"{o}dinger equation (NLSE). The presence of 
nonlinear interaction \cite{Nonlinearity}, in both the GPE and NLSE, has been exploited in, and is required for, the interpretation of dispersive shock waves as a train of grey solitons 
\cite{Engels_PRL_2008,Engels2009}. 
However, as we show below, qualitatively similar density modulations can form in a noninteracting case which does not support solitons. These incongruencies imply that the understanding of dispersive shock waves requires reassessment, including clarification of the role of actual quantum and thermal fluctuations which require beyond-mean-field descriptions. 

In this Letter, we study dispersive shock waves in a one-dimensional (1D) Bose gas, described by the Lieb-Liniger model \cite{ll63,*Lieb_II}, and show that the microscopic mechanism behind the formation of the oscillatory wave-train is quantum mechanical interference: the wavepacket that makes up a local density bump self-interferes with its own background upon expanding into it. This is in contrast to  a Gaussian wavepacket expanding into free space, which maintains its shape. Our results span the entire range of interaction strengths, from the noninteracting (ideal)  Bose gas regime, through the weakly-interacting or Gross-Pitaevskii regime, to the regime of infinitely strong interactions corresponding to the Tonks-Girardeau (TG) gas of hard-core bosons \cite{Girardeau1960,Girardeau_Bose_Fermi}. In all regimes, the interference contrast decreases with the reduction of the local phase coherence length. Moreover, in the weakly interacting regime, where the interference contrast is typically high, we show that thermal and quantum fluctuations can dramatically reduce the contrast due to shot-to-shot fluctuations in the position of interference fringes. 

\begin{figure}[tbp]
\includegraphics[width=8.3cm]{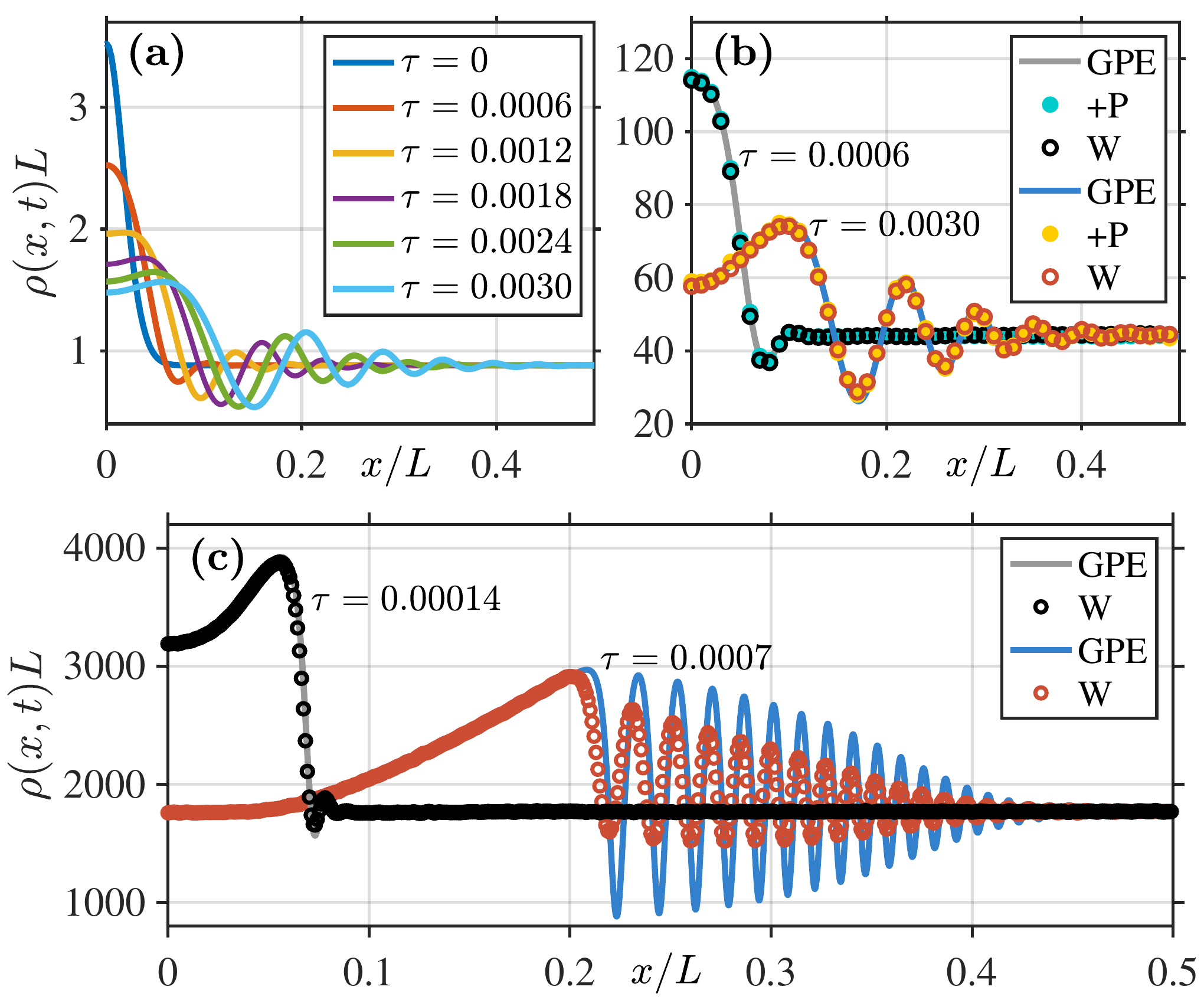}
\caption{Dispersive shock waves in an ideal and weakly interacting 1D Bose gas at $T=0$. In (a), we show the single-particle probability densities $\rho(x,t)\!=\!|\Psi(x,t)|^2$ for the ideal gas at different dimensionless times $\tau$, for $\beta\!=\!1$ and $\sigma/L\!=\!0.02$. Due to the reflectional symmetry about the origin, we only show the densities for $x\!>\!0$. In (b) and (c), we show the density profiles in the weakly interacting regime at two instances of time, with the GPE simulations represented by solid (grey and blue) lines. We also show the results of stochastic phase-space simulations \cite{errors}
using the truncated Wigner ($W$) and positive-$P$ ($+P$) approaches, which incorporate the effects of vacuum fluctuations (see text). The shape of $\rho(x,0)$ (not shown) in (b) and (c) is the same as in (a), except that it is now normalized to $N$: (b) $N\!=\!50$, $\gamma_{\mathrm{bg}}\!=\!0.1$; (c) $N\!=\!2000$,  $\gamma_{\mathrm{bg}}\!=\!0.01$ \cite{Comment3}. The dimensionless healing length $l_{\mathrm{h}}/L\!=\!1/\sqrt{\gamma_{\mathrm{bg}}}N_{{\mathrm{bg}}}$  is $l_{\mathrm{h}}/L\!\simeq \!0.072$ in (b), and $l_{\mathrm{h}}/L\!\simeq \!0.0057$ in (c), which can be compared to $\sigma/L\!=\!0.02$.}
\label{fig:single_particle_and_GPE}
\end{figure}

\emph{Ideal Bose gas.}---We begin our analysis with the simplest case of an ideal Bose gas. For analytical insight, we consider the initial wavefunction $\Psi(x,0)\!=\!\psi_{\mathrm{bg}}(1\!+\!\beta e^{-x^2/2\sigma^2})$, prepared prior to time $t\!=\!0$ as the ground state of a suitably chosen dimple potential, which subsequently evolves in a uniform potential of length $L$ with periodic boundary conditions.
The Gaussian bump has a width $\sigma$ and amplitude $\beta$ above a (real) constant background $\psi_{\mathrm{bg}}$
which fixes the normalization of the wavefunction to unity in the single-particle case, or to the total number of particles $N$ in the system 
\cite{Shape}. The dimensionless density $\overline{\rho}_{\mathrm{bg}}\!=\!\rho_{\mathrm{bg}}L\!=\!|\psi_{\mathrm{bg}}|^2L\!=\!N_{\mathrm{bg}}$ gives the number of particles in the background, with 
$N\!=\!N_{\mathrm{bg}}\big( 1+ \frac{\sqrt{\pi} \beta \sigma}{L} [ \beta\, \mathrm{erf}(\frac{L}{2\sigma}) +2 \sqrt{2}  \,\mathrm{erf} (\frac{L}{2\sqrt{2}\sigma}   )  ] \big)$. The wavefunction $\Psi(x,0)$ evolves according to the time-dependent Schr\"{o}dinger equation, whose solution can be written as 
\begin{equation}
\Psi(x,t)=\psi_{\mathrm{bg}}\big(1+\tfrac{\beta\sigma}{\sqrt{\sigma^2+i\hbar t/m}} e^{-x^2/2(\sigma^2+i\hbar t/m)}\big).
\label{initial_wf}
\end{equation} 

The corresponding density profile $\rho(x,t)\!=|\Psi(x,t)|^2$ is shown in Fig.~\ref{fig:single_particle_and_GPE}\,(a) at different dimensionless times $\tau\!=\!t/t_0$ (and before significant reflections off the boundary), where $t_0\!=\!mL^2/\hbar$ is the time scale, and $m$ is the mass of the particles. 
As we see, $\rho(x,t)$ displays all the known hallmarks of dispersive shock waves from the GPE (see below). In particular, the shock wave oscillations are chirped, with high-frequency and small-amplitude components located at the shock front. The wavefunction (\ref{initial_wf}) can be rewritten as $\Psi(x,t)\!=\!\psi_{\mathrm{bg}}[1\!+\!B(x,t)e^{i\varphi(x,t)}]$, so that the density $\rho(x,t)\!=\!|\Psi(x,t)|^2$ acquires a textbook form of quantum mechanical interference, $\rho(x,t)\!=\!\psi_{\mathrm{bg}}^2[1\!+\!B(x,t)^2\!+\!2B(x,t)\cos\varphi(x,t)]$, with the amplitude $B(x,t)\!\equiv\! \frac{\beta \sigma} {[\sigma^4+\hbar^2 t^2/m^2]^{1/4}}e^{-x^2\sigma^2/2[\sigma^4+\hbar^2 t^2/m^2]}$ and phase $\varphi(x,t)\!\equiv\!\frac{\hbar t x^2}{2m[\sigma^4+\hbar^2 t^2/m^2]}\!-\!\frac{1}{2}\mathrm{atan}\,\big(\frac{ \hbar^2 t^2}{m^2\sigma^4}\big)$. This means that the period of oscillations in the bulk of the shock train is $\sim\!\!2\sigma$ (with $\sigma$ being the only relevant length-scale in the problem), whereas the amplitude scales as $\propto \!\beta \sigma/ \sqrt{t}$.


\emph{Weakly interacting Bose gas in the GPE regime.}---We now move to consider repulsive pairwise delta-function interactions of strength $g$, and find ourselves in the realm of the Lieb-Liniger model \cite{ll63,o98,SupMat}. For a uniform system, the relevant dimensionless interaction parameter is $\gamma\!=\!mg/\hbar^2\rho$, where $\rho$ is the 1D density. For a nonuniform gas with a local density bump, one can introduce a local interaction parameter $\gamma(x)\!=\!mg/\hbar^2\rho(x)$ and use, e.g., the background value $\gamma_{\mathrm{bg}}\!=\!mg/\hbar^2\rho_{\mathrm{bg}}$ as the global interaction parameter to characterise the initial state (in addition to specifying the height and width of the bump). The weakly interacting regime of the Lieb-Liniger gas corresponds to $\gamma_{\mathrm{bg}}\! \ll \! 1\!$ [hence $\gamma(x)\!\ll \!1$ at any other $x$ within the bump], and the zero-temperature ($T\!=\!0$) dynamics of the system can be approximated by the GPE for the complex mean-field amplitude $\Psi(x,t)$:
\begin{equation}
i\hbar\partial_{t}\Psi(x,t)\!=\!\Big(\!-\tfrac{\hbar^2}{2m}\partial_{xx} + g|\Psi(x,t)|^2\Big)\Psi(x,t).
\end{equation} 

Dispersive shock waves forming under the GPE are shown in Fig.~\ref{fig:single_particle_and_GPE}\,(b) and (c), and are qualitatively similar to those in the ideal Bose gas. The interfering nature of the density ripples in this regime, where we no longer have an explicit analytic solution, can be revealed \cite{SupMat} via a wavelet transform known from signal processing theory \cite{debnath_book15a,WT-gravitational,colas16a,colas18a}. The only difference that arises here is that the interaction term in the GPE sets up a new lengthscale in the problem---the healing length $l_{\mathrm{h}}\!=\!\hbar/\sqrt{mg\rho_{\mathrm{bg}}}$ of the background. The healing length decreases with increasing interactions, and as soon as it becomes the shortest lengthscale in the problem (hence determining the effective UV momentum cutoff) it overtakes the role of $\sigma$ in determining the characteristic period of interference oscillations.  An example of this scenario is shown in Fig.~\ref{fig:single_particle_and_GPE}\,(c): here, the initial density profile is in the Thomas-Fermi regime (where the mean-field interaction energy per particle is much larger than the kinetic energy) with $l_{\mathrm{h}}\!<\!\sigma$, and the characteristic period of oscillations is $\sim\!2l_{\mathrm{h}}$. The trailing interference fringe of the shock-wave train propagates approximately at the speed of sound $c_s=\sqrt{g \rho_{\mathrm{bg}}/m}$ at the background density $\rho_{\mathrm{bg}}$, obtained from the Bogolibov spectrum of elementary excitations \cite{ll63,*Lieb_II}.

The GPE can be equivalently formulated in terms of superfluid hydrodynamics via Madelung's transformation to the density and phase variables, $\Psi(x,t)=\sqrt{\rho(x,t)}e^{i\phi(x,t)}$, and the velocity field $v(x,t)=\frac{\hbar}{m}\partial_x\phi(x,t)$, which yields
\begin{align}
\partial_t\rho& =-\partial_x(\rho v),
\label{eq:q-hydro-a}\\
\partial_t{v}& =-\partial_x\big(\tfrac{1}{2}v^2+\tfrac{g\rho}{m}-\tfrac{\hbar^{2}}{2m^{2}}\tfrac{1}{\sqrt{\rho}}\partial_{xx} \sqrt{\rho} \big).
\label{eq:q-hydro-b}
\end{align}
The last (dispersive) term in Eq. (\ref{eq:q-hydro-b}) is referred to as the quantum pressure term; it is this term that governs the formation of the oscillatory wave-train in the hydrodynamic approach.

The same quantum pressure term arises in the hydrodynamic-like formulation of the single-particle Schr\"{o}dinger equation after applying Madelung's transformation to the quantum mechanical wavefunction. This means that in the ideal Bose gas case, with $g\!=\!0$ in the above hydrodynamic equations, it is again the quantum pressure term that is responsible for producing dispersive shock wave oscillations in Fig.~\ref{fig:single_particle_and_GPE}\,(a). We note then, that the interaction $g$ is not necessary for the formation of the oscillatory shock wave train, and as such these oscillations cannot generally be interpreted as a train of grey solitons  \cite{Engels_PRL_2008,Engels2009} which do require the interactions to balance the wave dispersion. 

\begin{widetext}
\begin{figure*}[tbp]
\includegraphics[width=17.7cm]{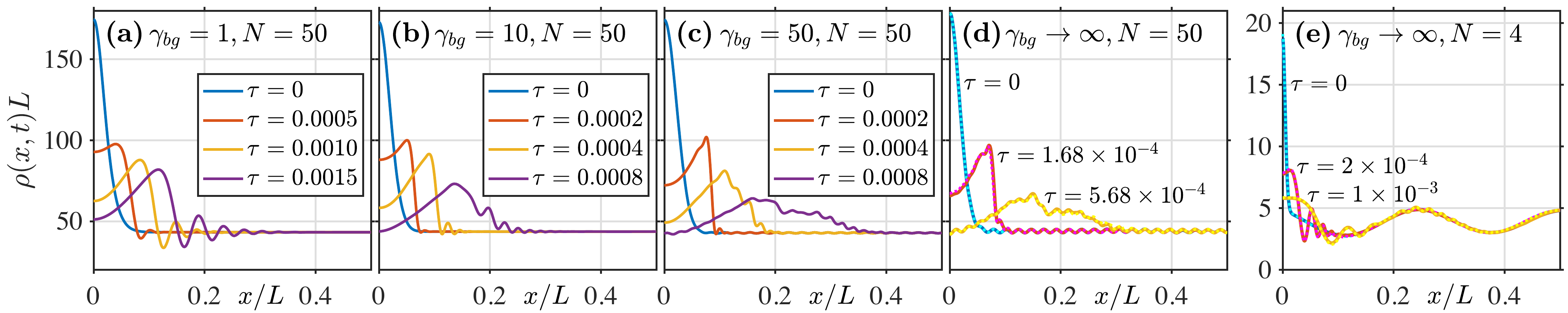}
\caption{Shock waves in a 1D Bose gas at $T=0$ for intermediate and strong interactions. In all panels, the iMPS simulation results are shown as full lines; in (d) and (e), the iMPS results are compared with exact diagonalization results (dotted lines), and show excellent agreement.
In (a)--(c), the trapping potential is chosen as $\bar{V}({\xi})\!=-\!\bar{V}_{bg}[1+\beta \exp(-\xi^2/2\bar{\sigma}^2)]^2$, with $\bar{\sigma}\!=\!0.02$ in all cases, and: (a) $\beta\!=\!0.98$, $\bar{V}_{bg}\!=\!1849$ (resulting in $N_{bg}\!\simeq \!43.2$); (b) $\beta\!=\!0.7$, $\bar{V}_{bg}\!=\!18705$ ($N_{bg}\!\simeq \!43.6$); (c) $\beta\!=\!0.38$, $\bar{V}_{bg}\!=\!92450$ ($N_{bg}\!\simeq \!43)$. Here, $\bar{V}(\xi)\!\equiv\! V(x)/E_0$, with $E_0\!=\!\hbar^2/mL^2$ being the energy scale. In (d), the trapping potential is chosen according to Eq.~(S33) of \cite{SupMat}, with $N_{bg}=44.03$, $\beta=1$, and $\bar{\sigma}=0.02$; in the Thomas-Fermi approximation, this would produce exactly the same initial density profile as in Fig.~\ref{fig:single_particle_and_GPE}\,(a), which, however, would not display the Friedel oscillations seen here. In (e), the trapping potential has the same shape as the one used for producing the ideal Bose gas initial density profile of Fig.~\ref{fig:single_particle_and_GPE}\,(a), except normalized to $N\!=\!4$.}
\label{fig:iMPS}
\end{figure*}
\end{widetext}

\emph{Strongly interacting and Tonks-Girardeau regimes.}---We now extend our analysis to increasingly stronger interaction strengths, from $\gamma_{\mathrm{bg}}\!\sim\!1$ to the TG limit of $\gamma_{\mathrm{bg}}\!\rightarrow \!\infty$ \cite{Girardeau1960,Girardeau_Bose_Fermi,ll63}. The shock wave dynamics in this regime are shown in Fig.~\ref{fig:iMPS} and are simulated using infinite matrix product states (iMPS) \cite{SupMat}, starting from the ground state of a dimple potential $V(x)$. The key observation here is that the amplitude of shock wave oscillations (i.e., the interference contrast) goes down with increasing $\gamma_{\mathrm{bg}}$ due to the reduction of the local phase coherence length of the gas. The phase coherence length of a 1D Bose gas crosses over from essentially the size of the system in the GPE regime down to the mean interparticle separation $1/\rho_{\mathrm{bg}}$ in the limit of $\gamma_{\mathrm{bg}}\!\gg \!1$ \cite{Cazalilla:2004}. Furthermore, for $\gamma_{\mathrm{bg}}\!\gg\!1$, the initial ground state density profile exhibits small-amplitude Friedel oscillations \cite{Friedel}, with a characteristic period equal to the mean interparticle separation $1/\rho_{\mathrm{bg}}$. This means that discerning between the deformations of these pre-existing oscillations and shock wave interference fringes, which form dynamically, becomes ambitious especially when the width $\sigma$ is on the same order of magnitude as $1/\rho_{\mathrm{bg}}$.


These observations become more evident in the TG limit of $\gamma_{\mathrm{bg}}\!\rightarrow \!\infty$,  where the mean interparticle separation in the background becomes the shortest lengthscale in the problem, related to the Fermi wavelength $\lambda_F\!=\!2\pi/k_F$ (with $k_F\!=\!\pi \rho_{\mathrm{bg}}$ being the Fermi wavevector at the background density) via $1/\rho_{\mathrm{bg}}\!=\!\lambda_F/2$.
Examples of evolving density profiles in the TG limit, obtained using exact diagonalization of a free fermion Hamiltonian and iMPS simulations \cite{SupMat}, are shown in Fig.~\ref{fig:iMPS} (d) and (e) for a relatively wide and a very narrow density bump. As we see in panel (d), dispersive shock wave oscillations in the TG gas do not form \cite{Tonks-absence} when the width of the bump $\sigma$ is larger than the phase coherence length $1/\rho_{\mathrm{bg}}$. The small density ripples seen in this case are simply evolving deformations of the initial Friedel oscillations \cite{MGPE}. When, however, $\sigma\!<\!1/\rho_{\mathrm{bg}}$, as in panel (e), we do observe small-scale dispersive shock waves, confined initially within a single Friedel-oscillation period of $1/\rho_{\mathrm{bg}}\!=\!\lambda_F/2$. The characteristic period of interference fringes in this case is determined by $\sigma$ as it is now the shortest lengthscale in the problem.

\emph{The effects of thermal and vacuum fluctuations.}---In order to understand the effect of thermal fluctuations on dispersive shock waves, we consider the finite temperature quasi-condensate regime of the 1D Bose gas \cite{petrov2000,kgds03,kgds05,Bouchoule-mom-corr,jabkb11,bsdk16}. This regime still corresponds to weak interactions, $\gamma_{bg}\!\ll \!1$, but we focus on temperatures of the initial thermal state lying within $\gamma_{\mathrm{bg}} \!\lesssim \!\overline{\mathcal{T}}\!\lesssim\!\sqrt{\gamma_{\mathrm{bg}}}$ \cite{kgds03,kgds05}, where $\overline{\mathcal{T}}\!=\!T/T_d$ is the dimensionless temperature, $T_d\!=\!\hbar^2\rho_{\mathrm{bg}}^2/2mk_B$ is the temperature of quantum degeneracy of the gas at density $\rho_{\mathrm{bg}}$, and $k_B$ is the Boltzmann constant \cite{Temperature}. In this range of temperatures, which are most readily accessible in ultracold atom experiments \cite{vvwkv08,jabkb11,bvw09}, the density-density correlations of the gas are dominated by thermal rather than vacuum fluctuations \cite{kgds03,kgds05}. Accordingly, the shock wave dynamics can be simulated using $c$-field techniques \cite{Castin:2000,Davis:2001b,bbdbg08}, which involve preparation of the initial thermal equilibrium state using the stochastic projected Gross-Pitaevskii equation (SPGPE) and subsequent real-time evolution according to the GPE.

\begin{figure}[tbp]
\includegraphics[width=8.2cm]{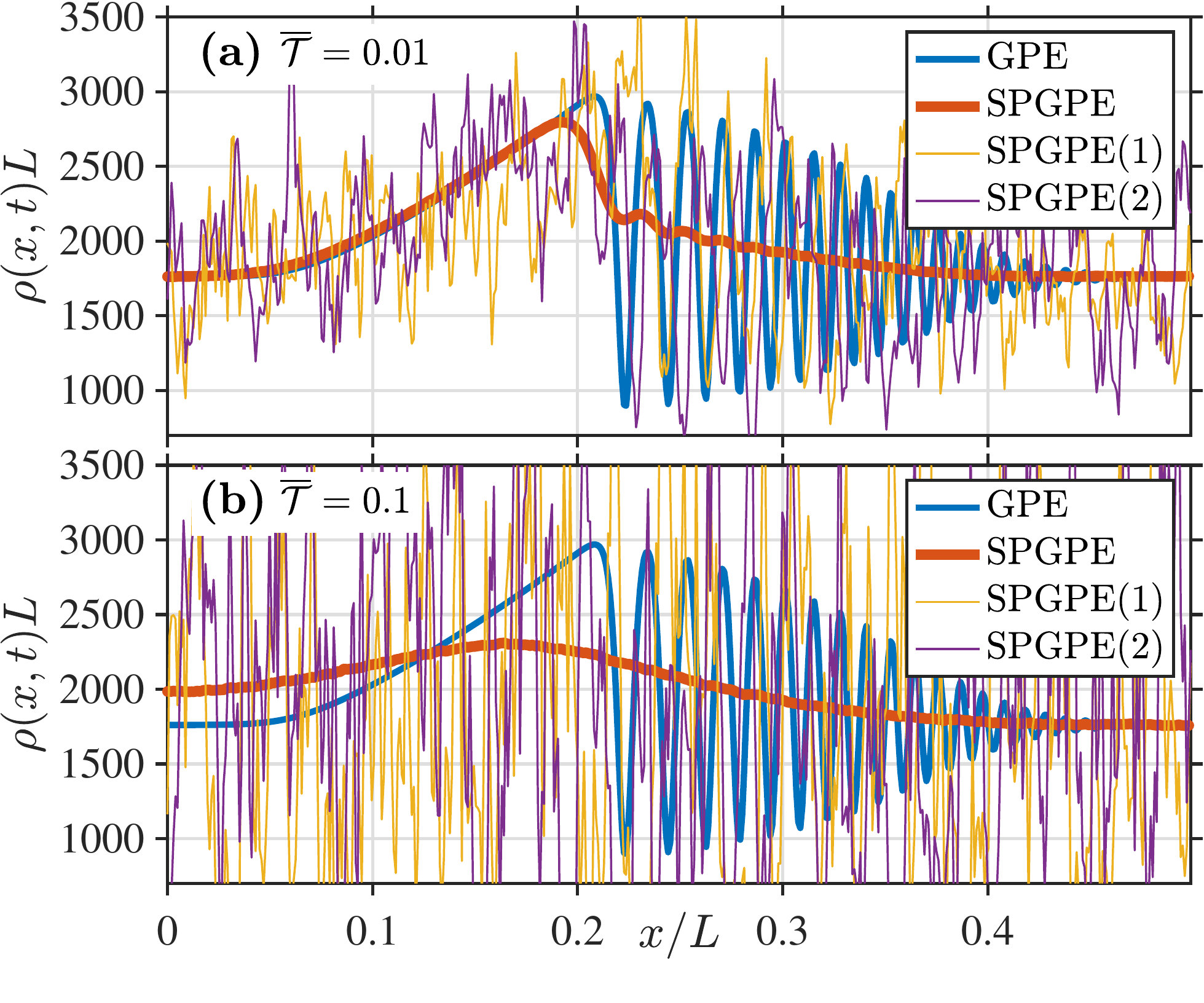}
\caption{Shock waves in a finite-temperature quasicondensate from SPGPE simulations (thick red lines). Shown are the final-time ($\tau\!=\!0.0007$) density distributions for $x\!>\!0$ and two different initial dimensionless temperatures: (a) $\overline{\mathcal{T}}\!=\!0.01$; and (b) $\overline{\mathcal{T}}\!=\!0.1$ \cite{Temperature}. Other parameters are as in Fig.~\ref{fig:single_particle_and_GPE}\,(c); the GPE results are shown here again as blue lines for comparison. The dimensionless thermal phase coherence length here can be expressed as $l_{T}/L=2/(\overline{\mathcal{T}}N_{\mathrm{bg}})$, giving: (a) $l_{T}/L\!\simeq \!0.1$; (b) $l_{T}/L\!\simeq \!0.01$. The SPGPE mean densities are the averages over 100,000 stochastic trajectories, whereas the thin lines show two sample trajectories.
}
\label{fig:finiteT}
\end{figure}

Examples of SPGPE simulations are shown in Fig.~\ref{fig:finiteT} for the same parameters as in Fig.~\ref{fig:single_particle_and_GPE}\,(c), but for two nonzero temperatures. As expected, the interference contrast is significantly reduced (compared to GPE results) due to thermal fluctuations and the resulting loss of phase coherence. Indeed, in the quasicondensate regime with density $\rho_{\mathrm{bg}}$, the thermal phase coherence length is given by $l_T\!=\!\hbar^2\rho_{\mathrm{bg}}/mk_BT$. From this estimate one can expect that the self-interfering shock wave train would lose its contrast when  $l_T$ becomes on the order of or smaller than the width of the bump $\sigma$, with oscillations eventually disappearing at sufficiently high temperatures. This is indeed what we see in Fig.~\ref{fig:finiteT}. However, the interference contrast in the example of Fig.~\ref{fig:finiteT}\,(a) is essentially lost at temperatures for which $l_T$ is still larger than $\sigma$; this can be explained by the shot-to-shot fluctuations in the position of interference fringes due to the same thermal fluctuations. Indeed, as can be seen from samples of individual stochastic SPGPE trajectories (shown as thin lines), even though these individual trajectories show high-contrast interference fringes (albeit with stochastic noise also present), the overall ensemble average over thousands of SPGPE realisations shows much lower interference contrast. This observation is consistent with the interpretation of the individual SPGPE trajectories representing individual experimental runs \cite{bbdbg08}, whereas the mean density corresponds to the ensemble average over many runs.

Finally, we consider the effect of quantum fluctuations on the shock wave interference contrast in the weekly interacting regime at $T\!=\!0$. These are treated using two stochastic phase-space methods, the truncated Wigner and positive-$P$ approaches \cite{Steel_1998}, and the iMPS method. The stochastic simulation results are shown in Fig.~\ref{fig:single_particle_and_GPE}\,(b) and (c), and are directly comparable to those based on the mean-field GPE. For the parameters of Fig.~\ref{fig:single_particle_and_GPE}\,(b) (very weak interactions), the truncated Wigner and positive-$P$ results agree with each other (in addition to being in excellent agreement with iMPS results \cite{SupMat}) within the respective error bars, and are close to the GPE results. In this regime, the quantum fluctuations have a negligible effect on the mean density and the interference contrast. For the parameters of Fig.~\ref{fig:single_particle_and_GPE}\,(c), on the other hand, the interactions are stronger (with the period of shock-wave oscillations determined by $l_{\mathrm{h}}$ rather than $\sigma$) and the quantum fluctuations have a more profound effect: the interference contrast is visibly reduced compared to the GPE prediction \cite{Intractable}. This is similar to the effect of thermal fluctuations discussed above, and can be attributed to shot-to-shot fluctuations in the position of interference fringes around the mean.

\emph{Conclusions.}---We have shown that the mechanism of formation of dispersive shock wave trains in a 1D Bose gas is quantum interference: the local perturbation self-interferes with its own background upon expanding into it. The interference contrast in this picture goes down with the reduction of the phase coherence length of the gas, and the picture holds true for all interaction strengths. We have also shown that thermal and quantum fluctuations can reduce the interference contrast further due to shot-to-shot fluctuations in the position of interference fringes around the mean. 
In the TG limit of infinitely strong interactions, where the phase coherence length is the same as the mean interparticle separation, the shock wave oscillations are absent for a sufficiently wide density bump (wider than the mean interparticle separation).
Apart from explaining the origin of density ripples in dispersive quantum shock waves, our results may serve as a test bed for new theoretical and computational techniques for many-body dynamics, such as the generalized hydrodynamics \cite{Doyon2017,GHD_1,GHD_2,Dubessy_2020}, and may shed new light on the understanding of dispersive shock waves in a variety of other contexts, such as in electronic systems described by the Calogero-Sutherland model and Korteweg-de Vries equations \cite{Abanov2006,Abanov_PRA_2012,Hoefger2016_review,Abanov2006}, or superfluids with higher-order dispersion \cite{mossman2020shock}.

K.\,V.\,K. acknowledges stimulating discussions with A.~G.~Abanov, V.~V.~Cheianov, J.~F.~Corney, M.~J.~Davis, and D.~M.~Gangardt. This work was supported through Australian Research Council (ARC) Discovery Project Grants No. DP170101423 and No. DP190101515, and by the ARC Centre of Excellence in Future Low-Energy Electronics Technologies (Project No. CE170100039).

\onecolumngrid
\newpage



\section*{Supplementary material}

tationally intractable due to large sampling errors.

# Supplemental Material for: What is a quantum shock wave?

S. A. Simmons,[1] F. A. Bayocboc, Jr.,[1] J. C. Pillay,[1] D. Colas,[1,2] I. P. McCulloch,[1] and K. V. Kheruntsyan[1]

[1]*School of Mathematics and Physics, University of Queensland, Brisbane, Queensland 4072, Australia*

[2]*ARC Centre of Excellence in Future Low-Energy Electronics Technologies,*
*University of Queensland, Brisbane, Queensland 4072, Australia*

(Dated: August 22, 2020)

## I. LIEB-LINIGER MODEL

The Lieb-Liniger Hamiltonian [S1] used to describe a trapped one-dimensional (1D) Bose gas of $N$ particles of mass $m$, with repulsive pairwise contact interaction between the particles, is given in second quantized form by

$$\hat{H} = \int dx \hat{\Psi}^\dagger \left( -\frac{\hbar^2}{2m}\frac{\partial^2}{\partial x^2} + V(x) \right) \hat{\Psi} + \frac{g}{2} \int dx \hat{\Psi}^\dagger \hat{\Psi}^\dagger \hat{\Psi} \hat{\Psi},$$
(S1)

where $\hat{\Psi}^\dagger(x)$ and $\hat{\Psi}(x)$ are the bosonic field creation and annihilation operators. The external trapping potential and the interparticle interaction strength are denoted by $V(x)$ and $g$, respectively. For harmonic transverse confinement with frequency $\omega_\perp$, which must be much stronger than the longitudinal confinements $V(x)$ in order to enable the reduction of a realistic three-dimensional (3D) system to an effective 1D model with frozen transverse degrees of freedom, the interaction strength can be approximated by $g \simeq 2\hbar\omega_\perp a$ away from confinement induced resonances [S2]. Here, $a$ is the 3D $s$-wave scattering length which is positive for repulsive interactions. In this Letter, we have considered evolution of the 1D Bose gas in a periodic box of length $L$ with uniform potential $V(x) = 0$. We write the external potential explicitly in Eq. (S1) so as to later consider the trap $V(x)$ which is required to prepare the desired initial density distributions as the ground state, before evolution in the uniform potential takes place after a sudden trap quench at time $t = 0$ to $V(x) = 0$.

At zero temperature ($T = 0$), the ground-state properties of a uniform Lieb-Liniger gas in the thermodynamic limit can be completely characterized by just a single dimensionless interaction parameter $\gamma = mg/\hbar^2\rho$ [S1, S3], where $\rho$ is the 1D (linear) density. For describing a nonuniform system, with the 1D density profile $\rho(x)$, one can instead use a local dimensionless interaction parameter $\gamma(x) = mg/\hbar^2\rho(x)$ [S4]. In the main text, the value of $\gamma(x)$ at the background density $\rho_{bg}$ is denoted via $\gamma_{bg}$.

At non-zero temperatures, a further dimensionless temperature parameter is required for characterising the Lieb-Liniger gas. This can be chosen to be defined relative to a characteristic temperature scale in the system (such as the temperature of quantum degeneracy $T_d = \hbar^2\rho^2/2mk_B$ for a uniform system), or in terms of $k_B T$ relative to a characteristic energy scale in the problem (such as $E_b = mg^2/2\hbar^2$, which is equivalent to the binding energy of a two-particle bound state that exists in the attractive counterpart of the model, with $g < 0$) [S4, S5]. The numerical study carried out in this work though will always be for a finite-size system of length $L$ and hence

there is an additional energy scale $\hbar^2/mL^2$ at our disposal, which we will use as necessary.

## II. MEAN-FIELD GPE AND SUPERFLUID HYDRODYNAMICS

In the weakly interacting limit, $\gamma_{bg} \ll 1$, we employ the mean-field approximation and set $\hat{\Psi}(x,t) \rightarrow \langle \hat{\Psi}(x,t) \rangle = \Psi(x,t)$, where $\Psi(x,t)$ is to be understood as the order parameter or the mean field amplitude of the system in the spontaneously broken symmetry approach. Making such an approximation in the Heisenberg equation of motion for the Bose annihilation operator results in the well-known mean-field Gross-Pitaevski equation (GPE), describing the zero-temperature dynamics of the gas:

$$i\hbar\frac{\partial\Psi(x,t)}{\partial t} = \left( -\frac{\hbar^2}{2m}\frac{\partial^2}{\partial x^2} + V(x) + g|\Psi(x,t)|^2 \right)\Psi(x,t).$$
(S2)

Here, $V(x)$ is the trapping potential which we set to zero during the evolution, but we include it here to indicate its relevance for the preparation of a desired initial density profile as the ground state of that potential.

In dimensionless units, the GPE takes the form

$$i\frac{\partial\bar{\Psi}(\xi,\tau)}{\partial\tau} = \left( -\frac{1}{2}\frac{\partial^2}{\partial\xi^2} + \bar{V}(\xi) + \bar{g}|\bar{\Psi}(\xi,\tau)|^2 \right)\bar{\Psi}(\xi,\tau),$$
(S3)

where we have used the dimensionless quantities $\xi = x/L$, $\tau = t/t_0$, $\bar{\Psi} = \Psi\sqrt{L}$, $\bar{V} = V/E_0$, and $\bar{g} = g/E_0 L = gmL^2/\hbar^2 = \gamma_{bg}N_{bg}$, with the spatial size of the system $L$ being used as the length scale, $t_0 = mL^2/\hbar$ the timescale, and $E_0 = \hbar/t_0 = \hbar^2/mL^2$ the energy scale.

If one considers the ideal (noninteracting) gas case, where there are no interactions, *i.e.*, $\bar{g} = 0$, then Eq. (S3) reduces to the dimensionless Schrödinger equation, where $\bar{\Psi}(\xi,\tau)$ is to be understood as the single-particle quantum mechanical wavefunction.

Converting to density and phase variables via Madelung's transformation, $\Psi(x,t) = \sqrt{\rho(x,t)}e^{i\phi(x,t)}$, Eq. (S3) results in the dimensionless superfluid hydrodynamic equations,

$$\frac{\partial\bar{\rho}}{\partial\tau} = -\frac{\partial}{\partial\xi}(\bar{\rho}\bar{v}),$$
(S4)

$$\frac{\partial\bar{v}}{\partial\tau} = -\frac{\partial}{\partial\xi}\left( \frac{1}{2}\bar{v}^2 + \bar{V}(\xi) + \bar{g}\bar{\rho} - \frac{1}{2}\frac{1}{\sqrt{\bar{\rho}}}\frac{\partial^2\sqrt{\bar{\rho}}}{\partial\xi^2} \right),$$
(S5)

where $\bar{\rho} = \rho L$ is the dimensionless density and $\bar{v} = vt_0/L = vmL/\hbar = \partial\phi/\partial\xi$ is the dimensionless velocity field.

We now consider the trapping potential that is required in order to prepare an initial wavefunction comprised of a Gaussian bump on a constant background,

$$\bar{\Psi}(\xi, 0) = \bar{\Psi}_{bg}\left(1 + \beta e^{-\xi^2/2\bar{\sigma}^2}\right) \tag{S6}$$

where $\bar{\sigma} = \sigma/L$ is the rms width of the Gaussian component. This results in an initial density profile given by

$$\bar{\rho}(\xi, 0) = N_{bg}\left(1 + \beta e^{-\xi^2/2\bar{\sigma}^2}\right)^2, \tag{S7}$$

with $N_{bg} = N\left(1 + \sqrt{\pi}\beta\bar{\sigma}[\beta\,\mathrm{erf}(\frac{1}{2\bar{\sigma}}) + 2\sqrt{2}\,\mathrm{erf}(\frac{1}{2\sqrt{2}\bar{\sigma}})]\right)^{-1}$, which can be obtained by requiring that the density normalises to the total number of particles in the system $N = \int_{-L/2}^{L/2} \rho(x)dx = \int_{-0.5}^{0.5} \bar{\rho}(\xi)d\xi$.

In order to find a trap potential $\bar{V}(\xi)$ which produces this density profile as its ground state we consider the stationary states of the GPE and hence require $i\frac{\partial\bar{\Psi}(\xi,\tau)}{\partial\tau} = \bar{E}\bar{\Psi}(\xi,\tau)$, where $\bar{E}$ is the dimensionless eigenenergy of the desired ground state. Rearranging Eq. (S3) for the trapping potential yields

$$\bar{V}(\xi) = \bar{E} + \frac{1}{2\bar{\Psi}(\xi,0)}\frac{\partial^2\bar{\Psi}(\xi,0)}{\partial\xi^2} - \bar{g}|\bar{\Psi}(\xi,0)|^2. \tag{S8}$$

Since $\bar{E}$ is a constant and only provides an energy shift to the trap, it can be set to $\bar{E} = 0$. Substituting the desired initial density profile into Eq. (S8) gives the required trapping potential in the weakly interacting regime;

$$\bar{V}(\xi) = \frac{\beta}{2\bar{\sigma}^4}\left(\xi^2 - \bar{\sigma}^2\right)\left[e^{\frac{\xi^2}{2\bar{\sigma}^2}} + \beta\right]^{-1} - \bar{g}\bar{\rho}(\xi, 0). \tag{S9}$$

We note that this solution also includes the trap required in the ideal gas regime, which can be found by simply setting $\bar{g} = 0$.

Examples of the trap potentials $\bar{V}(\xi)$ generating the ground state density profiles in Fig. 1 of the main text are shown here in Fig. S1.

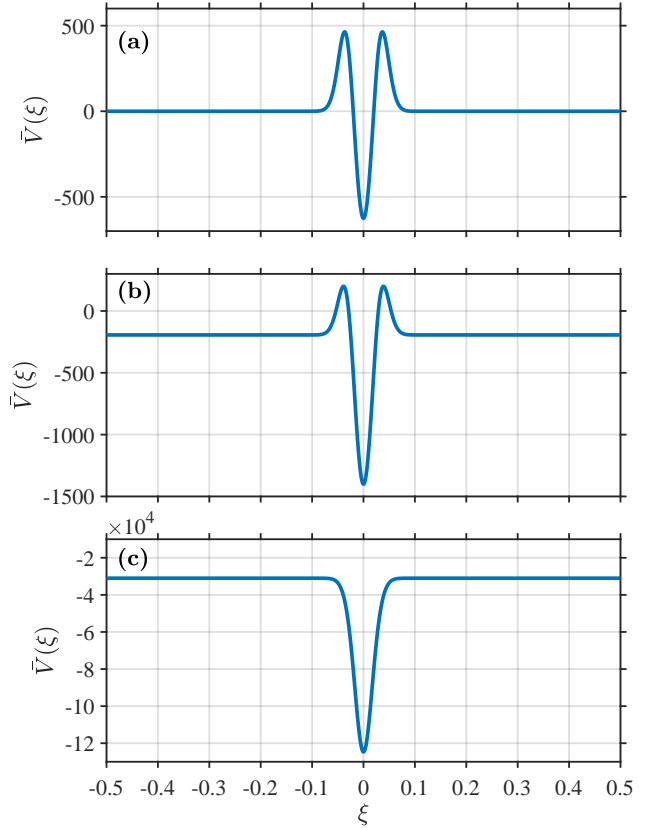

FIG. S1. External trapping potentials used to prepare the initial ground state density profiles in Fig. 1 of the main text. Panels (a), (b) and (c) here correspond, respectively, to the traps used in Figs. 1 (a), (b) and (c) of the main text.

## III. FINITE-TEMPERATURE QUASICONDENSATE REGIME

For studying the dynamics of a 1D quasicondensate, we use a classical or $c$-field technique [S6–S8], which is a proven approach for characterising the equilibrium and nonequilibrium properties of weakly interacting Bose gases at finite temperatures, particularly in 1D [S6, S8–S10]. The essence of the method is to treat the highly occupied modes of the quantum Bose field $\hat{\Psi}(x,t)$ as a classical field $\Psi_c(x,t)$, which satisfies the following simple growth stochastic projected Gross-Pitaevskii equation (SPGPE) for finding initial thermal equilibrium configuration:

$$d\Psi_c(\,x\,,t) = \mathcal{P}_c\left\{-\frac{i}{\hbar}\mathcal{L}_c\Psi_c(x,t)\,dt \right.$$
$$\left. + \frac{\Gamma}{k_BT}(\mu - \mathcal{L}_c)\Psi_c(x,t)\,dt + dW_\Gamma(x,t)\right\}. \tag{S10}$$

Here, $\mathcal{P}_c$ is the projection operator which sets up the high-energy cutoff for the classical field region, $\mathcal{L}_c$ is a shortcut for the Gross-Pitaevskii operator

$$\mathcal{L}_c = -\frac{\hbar^2}{2m}\frac{\partial^2}{\partial x^2} + V(x) + g|\Psi_c(x,t)|^2, \tag{S11}$$

$\mu$ is the chemical potential, and $T$ is the temperature of the reservoir of low-occupancy modes (treated as static) to which the $c$-field is coupled. In addition, $\Gamma$ is the growth rate, whereas the last term $dW_\Gamma(x,t)$ is the associated complex white noise, with the following nonzero correlation:

$$\langle dW_\Gamma^*(x,t)dW_\Gamma(x',t)\rangle_{\mathrm{stoch}} = 2\Gamma\delta(x-x')dt. \tag{S12}$$

Here and hereafter, $\langle...\rangle_{\mathrm{stoch}}$ refers to stochastic averaging over a sufficiently large number of stochastic trajectories.

Integrating each stochastic realisation of the SPGPE, initiated from a random initial noise, for a sufficiently long time (ergodicity assumed), is equivalent to sampling configurations from a canonically distributed Gibbs ensemble. Expectation values of physical observables then correspond to ensemble averages over a large number of stochastic realisations of the SPGPE.

The numerical value of the growth rate $\Gamma$ in Eq. (S10) has no consequence for the equilibrium configurations, and hence

can be chosen for numerical convenience. Furthermore, while the inclusion of an energy cutoff is crucial in higher dimensions in order to prevent a divergence of the atomic density, its role is less crucial in 1D as the classical field predictions for the atomic density do not diverge, even in absence of an energy cutoff, and are quantitatively correct for degenerate gases. For this reason, our simulations were performed without explicitly imposing the projection operator $\mathcal{P}_c$, in which case the cutoff merely determined the size of the computational basis used for the numerical calculations, and the results did not strongly depend on the cutoff once it was chosen sufficiently large.

Using the dimensionless variables introduced earlier, in addition to defining a dimensionless growth rate $\bar{\Gamma} = \Gamma t_0$, dimensionless temperature $\bar{T} = k_B T / E_0$, and the dimensionless $c$-field $\bar{\Psi}_c(\xi, \tau) = \Psi_c(x, t)\sqrt{L}$, the SPGPE can be rewritten in the following dimensionless form:

$$d\bar{\Psi}_c(\xi, \tau) = \mathcal{P}_c\Big\{-i\bar{\mathcal{L}}_c\bar{\Psi}_c(\xi, \tau)\, d\tau$$
$$+ \frac{\bar{\Gamma}}{\bar{T}}(\bar{\mu} - \bar{\mathcal{L}}_c)\bar{\Psi}_c(\xi, \tau)\, d\tau + d\bar{W}_{\bar{\Gamma}}(\xi, \tau)\Big\} \quad (S13)$$

where

$$\bar{\mathcal{L}}_c = -\frac{1}{2}\frac{\partial^2}{\partial \xi^2} + \bar{V}(\xi) + \bar{g}|\bar{\Psi}_c(\xi, \tau)|^2, \quad (S14)$$

and

$$\langle d\bar{W}_{\bar{\Gamma}}^*(\xi, \tau)d\bar{W}_{\bar{\Gamma}}(\xi', \tau)\rangle_{\text{stoch}} = 2\bar{\Gamma}\delta(\xi - \xi')d\tau. \quad (S15)$$

The dimensionless temperature parameter $\mathcal{T}$ introduced in the main text (which characterises the temperature of the gas relative to the temperature of quantum degeneracy) and the dimensionless temperature $\bar{T}$ introduced here are related by: $\mathcal{T} = 2\bar{T}/\bar{\rho}_{\text{bg}}^2 = 2\bar{T}/N_{\text{bg}}^2$.

Once the SPGPE is evolved to its final thermal equilibrium configuration in a given trapping potential (which we chose to be the same as in Eqs. (S8) and (S9), with $\bar{\Psi}_c(\xi, 0)$ taking the role of $\bar{\Psi}(\xi, 0)$ as to provide the same initial density profiles as in the respective $T = 0$ examples), we then simulate the actual shock wave dynamics by evolving each stochastic realization of the $c$-field in real time according to the standard GPE, in other words, by omitting the growth and noise terms in Eq. (S13) and setting $\bar{V}(\xi) = 0$.

Examples of density profiles from SPGPE simulations for the parameters of Fig. 1 (b) of the main text are shown here in Fig. S2 for three different temperatures. As we see, the interference contrast reduces with increasing temperature which is due to the reduction of the thermal phase coherence length $l_T$ in the system.

## IV. TRUNCATED WIGNER AND POSITIVE-$P$ METHODS

In order to explore the effects of quantum fluctuations on dispersive shock waves, we turn to stochastic phase-space approaches, namely the truncated Wigner and positive-$P$ methods [S8, S11–S15]. In these approaches, quantum fluctuations

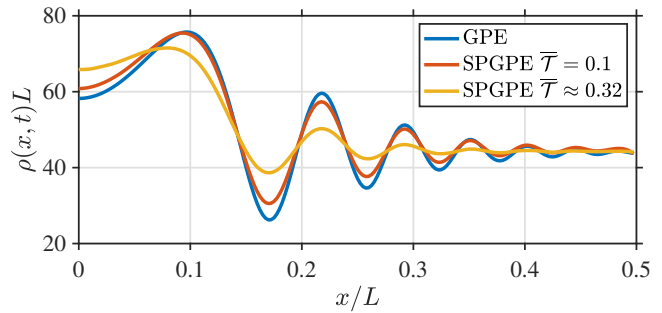

FIG. S2. Shock waves in the finite temperature quasicondensate regime from SPGPE simulations. Shown are the final-time ($\tau = 0.003$, with other parameters as in Fig. 1 (b) of the main text) density distributions for $x > 0$ and two different initial dimensionless temperatures $\bar{\mathcal{T}} = T/T_d$ as shown, where $T_d = \hbar^2\rho^2/2mk_B$ is evaluated at the background density $\rho_{\text{bg}}$. The zero-temperature GPE result is shown here again (blue line) for comparison. The thermal phase coherence lengths in the background, for $\bar{\mathcal{T}} = 0.1$ and $\bar{\mathcal{T}} = 0.32$, are $l_T/L \simeq 0.45$ and $l_T/L \simeq 0.15$, respectively, which can be compared with the width of the initial bump $\sigma/L = 0.02$.

are represented in phase space via stochastic distributions of $c$-field trajectories, and by taking ensemble averages over many trajectories, physical observables such as the real space density can be computed.

The truncated Wigner approach is an approximate phase-space method where the third- and higher-order derivative terms in the evolution equation for the Wigner quasiprobability distribution function are ignored (truncated). Such truncation renders the evolution equation as a true Fokker-Planck equation which can equivalently be formulated in terms of stochastic differential equations for complex $c$-fields. For the Lieb-Liniger Hamiltonian considered here, with $s$-wave scattering interactions, these differential equations have the same form as the standard time-dependent GPE. However, the quantum effects are incorporated via noise added to the initial state of the system, which is then sampled stochastically, with each realization evolving according to the GPE. In this work, we consider zero-temperature initial ground states with vacuum excitations in the Bogoliubov theory. In this case, each realization of the stochastic Wigner field $\psi_W(x, 0)$ is initialized as [S8]

$$\psi_W(x, 0) = \Psi_0(x, 0) + \sum_{j=1}^{M_B}\left[\eta_j u_j(x) + \eta_j^* v_j^*(x)\right], \quad (S16)$$

where $\eta_j$ is a complex Gaussian noise term of zero mean ($\langle\eta_j\rangle_{\text{stoch}} = 0$) and with correlations $\langle\eta_j^*\eta_k\rangle_{\text{stoch}} = \frac{1}{2}\delta_{jk}$, and index $j = 1, 2, ...M_B$ enumerates the Bogoliubov excitations with $u_j(x)$ and $v_j(x)$ being the respective amplitudes. We note that the noise correlations are the zero-temperature limit of a more general expression $\langle\eta_j^*\eta_k\rangle_{\text{stoch}} = (\bar{n}_j + \frac{1}{2})\delta_{jk}$, where $\bar{n}_j = \left[e^{\epsilon_j/k_B T} - 1\right]^{-1}$ are the Bose occupation numbers (that vanish at $T = 0$) of Bogoliubov modes with energies $\epsilon_j$ at temperature $T$ [S8, S12]. In addition, $\Psi_0(x, 0)$ represents the condensate mode wavefunction found as the ground state of the time-independent GPE. In dimensionless

form, it is determined from Eq. (S6), but renormalized to $N_0$ such that the total number of particles $N = N_0 + N_{nc}$ is the same as in the pure mean field approach of Sec. II, to which these truncated Wigner results are compared in the main text. Here, $N_0 = \int dx\, |\Psi_0(x,0)|^2$ is number of particles in the condensate mode, whereas $N_{nc} = \int_{-L/2}^{L/2} dx \sum_{j=1}^{M_B} |v_j(x)|^2$ is the expectation value of the population in the Bogoliubov modes (total number of non-condensate particles) at zero temperature.

The truncated Wigner approximation is valid when the total number of particles in the system $N$ greatly exceeds the number of relevant Bogoliubov modes $M_B$. All simulation results reported in this work using the truncated Wigner approach were carried out with the ratio $N/M_B \approx 5$, although variations around this value between $N/M_B \approx 2$ and $N/M_B \approx 10$ produced essentially the same results; the maximum difference between the densities corresponding to $N/M_B \approx 2$ and $N/M_B \approx 10$ was 6.4% for Fig. 1(b) [3.6% for Fig. 1(c)], and was located at the fringe minimum closest to the origin. We also recall that in the Wigner formalism, stochastic averages correspond to expectation values of symmetrically ordered operator products. Hence, the real space density is calculated as an average over many stochastic Wigner trajectories using $\rho(x,t) = \langle \psi_W^*(x,t)\psi_W(x,t)\rangle_{stoch} - \frac{1}{2}\delta_c(x,x)$, where $\frac{1}{2}\int_{-L/2}^{L/2} \delta_c(x,x)\,dx = \frac{1}{2}M_B$ represents the half quantum per Bogoliubov mode vacuum noise that is included in the Wigner formalism. On a computational grid of spacing $\Delta x = L/M$, where $M$ is the number of grid points, the projected delta function $\delta_c(x,x)$ is given by $\delta_c(x,x) = M_B/(M\Delta x)$ [S8]. In dimensionless form, the real-space density $\bar{\rho}(\xi,\tau) = \rho(x,t)L$ calculated from an average over Wigner trajectories can therefore be rewritten as $\bar{\rho}(\xi,\tau) = \langle \bar{\psi}_W^*(\xi,\tau)\bar{\psi}_W(\xi,\tau)\rangle_{stoch} - \frac{1}{2}M_B$, where $\bar{\psi}_W = \psi_W\sqrt{L}$.

Unlike the truncated Wigner approach, the positive-$P$ approach is exact (contingent on a vanishing 'boundary term' in the corresponding Fokker-Planck equation [S11, S16]) in the sense that no higher than second-order derivative terms arise in the corresponding Fokker-Planck equation for the Hamiltonian with $s$-wave scattering interactions; accordingly, no truncation is required in order to formulate the dynamics of the corresponding complex $c$-fields as stochastic differential equations. However, this method can become numerically unstable after relatively short times, when it develops large sampling errors due to the 'boundary term problem' [S17]. In the positive-$P$ approach, the dynamics of two independent complex $c$-fields $\psi(x,t)$ and $\tilde{\psi}(x,t)$, which correspond, respectively, to the field operators $\hat{\Psi}(x,t)$ and $\hat{\Psi}^\dagger(x,t)$, are simulated according to the following stochastic differential equations:

$$\frac{\partial \psi}{\partial t} = \frac{i}{\hbar}\left(\frac{\hbar^2}{2m}\frac{\partial^2}{\partial x^2} - V(x) - g\tilde{\psi}\psi\right)\psi + \sqrt{-i\frac{g}{\hbar}\psi^2}\,\zeta_1(x,t),$$
(S17)

$$\frac{\partial \tilde{\psi}}{\partial t} = -\frac{i}{\hbar}\left(\frac{\hbar^2}{2m}\frac{\partial^2}{\partial x^2} - V(x) - g\tilde{\psi}\psi\right)\tilde{\psi} + \sqrt{i\frac{g}{\hbar}\tilde{\psi}^2}\,\zeta_2(x,t).$$
(S18)

Here $\zeta_j(x,t)$ $(j = 1,2)$ are real, delta-correlated Gaussian noise terms that are generated dynamically and represent the vacuum fluctuations, with $\langle\zeta_j(x,t)\rangle_{stoch} = 0$ and $\langle\zeta_j(x,t)\zeta_k(x',t)\rangle_{stoch} = \delta_{jk}\delta(x-x')$. In this formulation, physical observables described by normally-ordered operator products can be computed from stochastic averages over direct products of $c$-fields, $e.g.$, $\rho(x,t) = \langle\hat{\Psi}^\dagger(x,t)\hat{\Psi}(x,t)\rangle = \langle\tilde{\psi}(x,t)\psi(x,t)\rangle_{stoch}$. In our simulations we have assumed that the gas is initially in a coherent state, and hence the initial conditions are taken to be $\psi(x,0) = \tilde{\psi}(x,0) = \Psi(x,0)$, where $\Psi(x,0)$ is the desired initial wavefunction, given in dimensionless form by Eq. (S6) and normalised to the total particle number $N$.

## V. WAVELET TRANSFORM

Unlike the Fourier transform that decomposes a signal using sine and cosine as basis functions, which are localized in Fourier space and delocalised in real space, the wavelet transform [S18] uses basis functions, called wavelets or merit functions, that are localized in both the real and Fourier spaces. The wavelet transform was originally developed in signal processing to obtain a representation of the signal in both time and frequency, as it is done $e.g.$, in the analysis of the gravitational-wave signal from a binary black hole merger [S19]. However, it has also been recently used to analyze interference phenomena between different wave packets [S18] or self-interference of a single wave packet [S20, S21] in the context of condensed matter physics, where the wavefunction, $i.e.$, the signal, is now represented in both position and momentum.

For a 1D wave packet, described by a dimensionless complex valued wavefunction $\bar{\Psi}(\xi,\tau)$ [or alternatively the mean-field amplitude in the GPE, or the $c$-field amplitude $\bar{\Psi}_c(\xi,\tau)$ in the SPGPE approach] in position space at time $\tau$, the dimensionless instantaneous wavelet transform $\mathbb{W}(\xi,q;\tau)$ is defined via

$$\mathbb{W}(\xi,q;\tau) = \frac{1}{\sqrt{|q|}}\int_{-0.5}^{+0.5}\bar{\Psi}(\xi',\tau)\mathcal{G}^*\left(\frac{\xi'-\xi}{q}\right)d\xi',$$
(S19)

Here, $\mathcal{G}$ is the wavelet, which we will choose to be the Gabor wavelet [S18]

$$\mathcal{G}(z) = \sqrt[4]{\pi}\exp(i\omega_\mathcal{G} z)\exp(-z^2/2^2),$$
(S20)

$i.e.$, a simple Gaussian envelope that has an internal phase $\omega_\mathcal{G}$ and that is square-integrable.

The scalogram or the map $|\mathbb{W}(\xi,q;\tau)|^2$ in the $(\xi,q)$-space at a given time $\tau$ gives the instantaneous cross-correlation between the wavefunction and the Gabor wavelet $\mathcal{G}$ and allows one to independently follow the propagation of different modes that compose the wave packet. Such a representation of a wavepacket is particularly useful for analysing chirped interfering signals, in which the interference at any position $x$ is revealed by the presence of two separate momentum components along $k$.

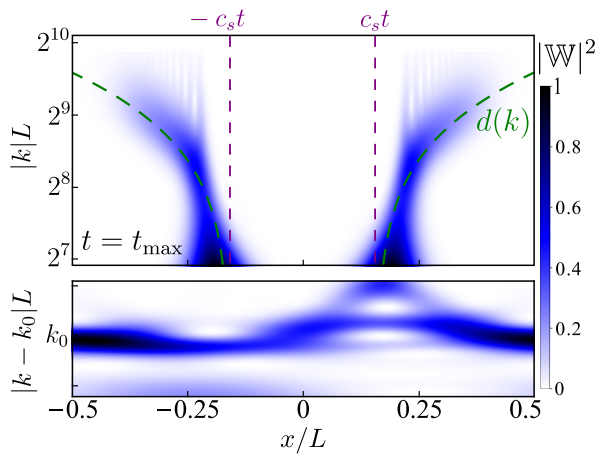

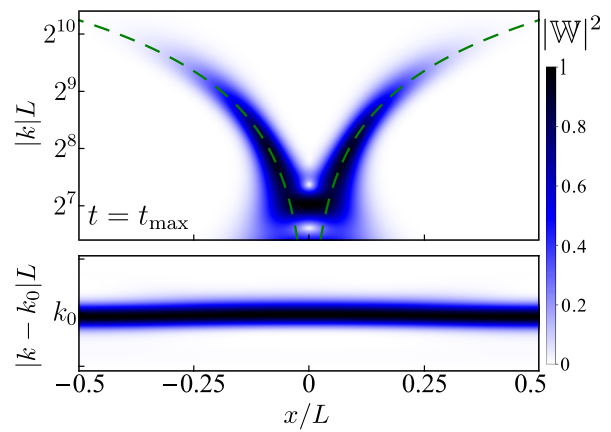

FIG. S3. Wavelet transform scalogram $|\mathbb{W}(x,k;t)|^2$ corresponding to the final-time ($\tau_{\max} = 0.0007$) GPE density profile of Fig. 1 (c) of the main text. As the wavelet transform is not well-defined at $k = 0$, we perform it using a small momentum shift $k_0$ in order to detect the background density component lying around zero momentum. The integration is done with $k_0 L = 2^6$ and $\omega_{\mathcal{G}} = 7$. The green dashed lines trail the displacements $d(k)$ derived from the Bogoliubov dispersion relation (see text), whereas the purple vertical dashed lines show the location of a test point (at time $\tau_{\max}$, starting from the origin) moving at the speed of sound at the background density, $c_s = \sqrt{g\rho_{\mathrm{bg}}/m}$.

FIG. S4. Wavelet transform scalogram $|\mathbb{W}(x,k;t)|^2$ corresponding to the single-particle wavefunction as in Fig. 1 (a) of the main text, except for $\sigma/L = 0.002$ and evaluated at time $\tau_{\max} = 0.0003$. The green dashed lines trail the displacements $d(k)$ derived from the free particle dispersion relation (see text). The background mode is obtained using the same method as in Fig. S3. The integration is done with $k_0 L = 2^5$ and $\omega_{\mathcal{G}} = 6$.

It is important to note that the wavelet transform obeys an uncertainty principle between its resolution in position ($\Delta_\xi$) and momentum ($\Delta_q$), so that the product $\Delta_\xi \Delta_q$ cannot be arbitrarily small. In the context of a quantum mechanical wavefunction as the signal, this product can be chosen to correspond to the Heisenberg uncertainty relation. The resolution in a given range ($\xi$, $q$) can then be numerically adapted to better visualize the region of interest, by using the wavelet frequency $\omega_{\mathcal{G}}$ or the signal sample rate.

An example of a scalogram, representing the shock wave train of Fig. 1 (c) of the main text at time $\tau = 0.0007$, is shown in Fig. S3 where we see two momentum branches (for both $x > 0$ and $x < 0$), or two distinct momentum components along $k$ for any given $x$ away from the central region near $x = 0$. The zero-momentum branch is identified with the uniform, delocalized background of the wave-packet, whereas the nonzero momentum branch corresponds to the local density bump, which interferes with the background as it expands into it. The nonzero branch trails a displacement $d(k) = v_{\mathrm{g}}(k)t$ of the wave-vector $k$, corresponding to the group velocity $v_{\mathrm{g}}(k) = \frac{1}{\hbar}\partial E_{\mathrm{B}}(k)/\partial k$, where $E_{\mathrm{B}}(k) = \sqrt{\frac{\hbar^2 k^2}{2m}\left(\frac{\hbar^2 k^2}{2m} + 2g\rho_{\mathrm{bg}}\right)}$ is the well-known dispersion relation for the Bogoliubov excitations. In dimensionless form, the displacement of Bogoliubov excitations takes the form $\bar{d}(q) = \bar{v}_{\mathrm{g}}(q)\tau$, where $q = kL$ is the dimensionless wave-number, $\bar{v}_{\mathrm{g}}(q) = \partial \bar{E}_{\mathrm{B}}(q)/\partial q$, and $\bar{E}_{\mathrm{B}}(q) = \sqrt{\frac{q^2}{2}\left(\frac{q^2}{2} + 2\bar{g}\bar{\rho}_{\mathrm{bg}}\right)} = \sqrt{\frac{q^2}{2}\left(\frac{q^2}{2} + 2\gamma_{\mathrm{bg}}N_{\mathrm{bg}}^2\right)}$, with $\bar{E}_{\mathrm{B}} \equiv E_{\mathrm{B}}/E_0$.

For comparison with the interacting case, in Fig. S4 we also show a wavelet transform scalogram for an ideal (noninteracting) Bose gas at zero temperature, which itself is equivalent to the solution of the single-particle Schrödinger equation, Eq. (1) of the main text. As previously, we see two momentum branches at any given $x$ away from the origin, which are the 'smoking gun' of interference occurring in the evolving density profile; the near horizontal branch around $k_0$ corresponds to the nonzero background, whereas the upper branches corresponds to the localized density bump. The displacement $d(k) = v_{\mathrm{g}}(k)t$ (shown by the dashed green line) corresponds to a test point moving at the group velocity $v_{\mathrm{g}}(k) = \frac{1}{\hbar}\partial E(k)/\partial k = \hbar k/m$, where $E(k) = \hbar^2 k^2/2m$ is the free particle parabolic dispersion. In dimensionless form, these convert to $\bar{d}(q) = \bar{v}_{\mathrm{g}}(q)\tau$, where $q = kL$, $\bar{v}_{\mathrm{g}}(q) = \partial \bar{E}(q)/\partial q |_{q=q}$, and $\bar{E}(q) = q^2/2$.

The same method can be applied to the shock waves obtained from the SPGPE simulations for finite temperature quasicondensates. The wavelet energy density $|\mathbb{W}(\xi,q;\tau)|^2$ is computed from the SPGPE complex $c$-field, for each individual stochastic trajectory, and then averaged. The scalograms corresponding to the cases in Fig. 3 of the main text, for temperatures $\overline{\mathcal{T}} = 0.01$ and $\overline{\mathcal{T}} = 0.1$, are shown here in Fig. S5 (a) and (b), respectively. At low temperature [$\overline{\mathcal{T}} = 0.01$, Fig. S5 (a)], there is background noise due to the thermal fluctuations, and the two distinct momentum branches—the signature of shock wave self-interference—are losing their visibility on the wavelet scalogram. At higher temperature [$\overline{\mathcal{T}} = 0.1$, Fig. S5 (b)], the two momentum branches that were once distinct can now barely be distinguished from the background noise. In this temperature regime, thermal fluctuations have lead to a significant loss of mean-field coherence, resulting in an almost complete loss of contrast of the interference fringes.

Similarly to the SPGPE case, the wavelet transform analysis can be applied to the stochastic field $\psi_W(x,t)$ in the truncated Wigner approach, wherein the wavelet energy density

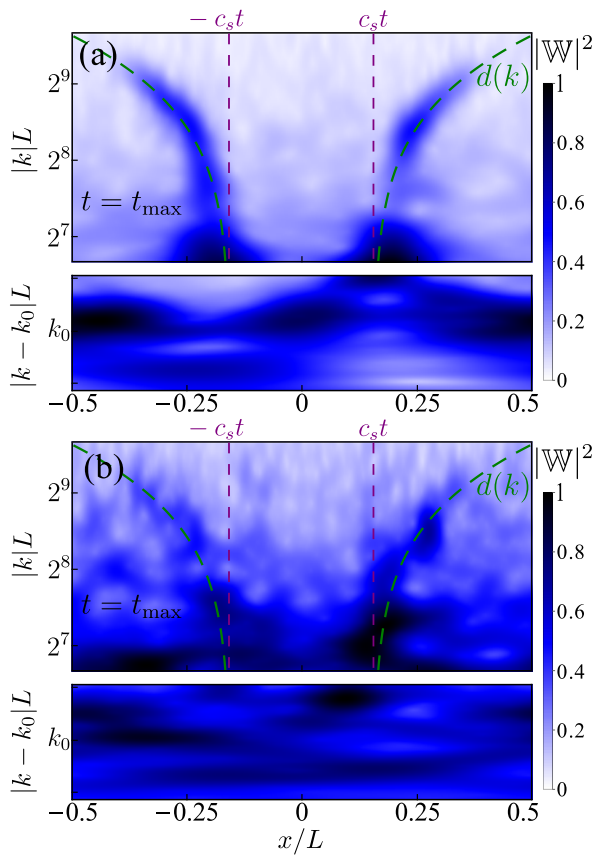

FIG. S5. Wavelet transform scalograms $|\mathbb{W}(x, k; t)|^2$ corresponding to the SPGPE simulation results of Fig. 3 (a) and (b) of the main text. The integration is done with $k_0 L = 2^6$ and $\omega_{\mathcal{G}} = 7$.

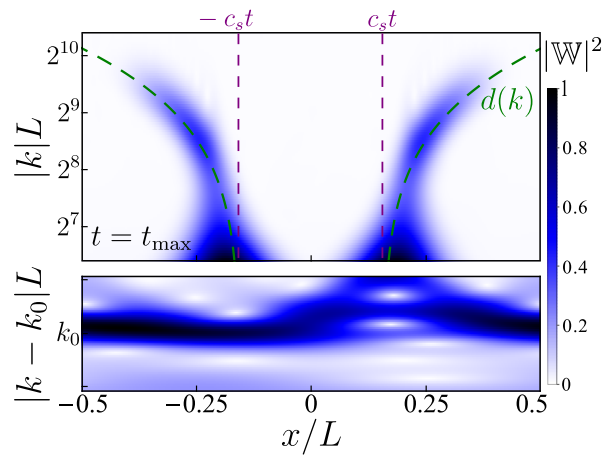

FIG. S6. Wavelet transform scalogram $|\mathbb{W}(x, k; t)|^2$ corresponding to the truncated Wigner simulation results of Fig. 1 (c) of the main text at $\tau_{\max} = 0.0007$. The integration is done with $k_0 L = 2^6$ and $\omega_{\mathcal{G}} = 7$.

$|\mathbb{W}(\xi, q; \tau)|^2$ is again computed for each individual stochastic trajectory and then averaged over many stochastic realisations. We recall that the stochastic nature of the field $\psi_W(x, t)$ now represents zero-temperature quantum fluctuations, rather than thermal fluctuations in the SPGPE approach. The resulting scalogram, for the parameter values of Fig. 1(c) of the main text at time $\tau_{\max} = 0.0007$, is shown here in Fig. S6 and can be directly compared to the scalogram of Fig. S3 from the mean-field GPE. As we see from this comparison, the two momentum branches are clearly identifiable in the truncated Winger wavelet transform, implying that the interfering nature of the shock wave oscillations is present even at the level of individual stochastic Wigner trajectories. This is despite the fact that the noisy nature of the signal due to quantum fluctuation may mask—to the naked eye—the interference oscillations in the individual trajectories. However, after averaging over many stochastic trajectories the two momentum branches in the scalogram become clearly identifiable.

## VI. MATRIX PRODUCT STATE METHODS

The Lieb-Liniger model Eq. (S1) can be treated with matrix product state (MPS) numerical methods [S22–S24], via

real-space discretization controlled by the lattice spacing $\Delta x$ [S25]. The simplest discretization procedure [S25–S27] amounts to replacing the field operator $\hat{\Psi}(x_j)$ by $\hat{b}_j/\sqrt{\Delta x}$, where $\hat{b}_j$ is the bosonic creation operator on site $j$ ($j = 1, 2, ...M$, for $M = L/\Delta x$ lattice sites) and $x_j = j\Delta x$. This discretization procedure leads to the Hamiltonian for the Bose-Hubbard model,

$$\hat{H} = -J \sum_j \left( \hat{b}_j^\dagger \hat{b}_{j+1} + \text{h.c.} \right) + \frac{U}{2} \sum_j \hat{b}_j^{\dagger 2} \hat{b}_j^2 + \sum_j V(x_j) \hat{b}_j^\dagger \hat{b}_j,$$
$$(S21)$$

where $J = \hbar^2/(2m\Delta x^2)$ and $U = g/\Delta x$. The validity of the discrete Bose-Hubbard model as an approximate description of the Lieb-Liniger model in the continuum limit $\Delta x \to 0$ has been discussed previously [S27–S29] and requires small average lattice occupancy, $\langle \hat{b}_j^\dagger \hat{b}_j \rangle \ll 1$, for ground state calculations, and hence $\Delta x \ll 1/\rho(x_j)$, where $\rho(x_j)$ is the local 1D density. For dynamical simulations, the lattice spacing has to also satisfy the condition $\langle \hat{b}_j^\dagger \hat{b}_j \rangle \ll 1/\gamma(x_j)$ [S28], which can become more important in the strongly interacting regime of $\gamma(x_j) \gg 1$.

In MPS methods, one needs to consider carefully the boundary conditions. Here, we used the infinite matrix product state (iMPS) method [S30], with a unit cell of $M$ sites, since this avoids large Friedel oscillations arising from vanishing boundary conditions (also known as open boundary conditions in the MPS community), and avoids the prohibitive increase in entanglement that comes with actual periodic boundary conditions (PBC) [S24]. This corresponds to an infinite periodic array of density bumps, spaced $M$ sites apart and produces a similar effect to periodic boundaries. As such, the iMPS method is directly comparable to exact diagonalization and other methods used in the main text with PBC, but differs in some respects; most notably the correlation functions always decay at long distance beyond the unit cell, whereas for PBC they are strictly periodic. Infinite boundary conditions (IBC) [S31] would also be a candidate method for these

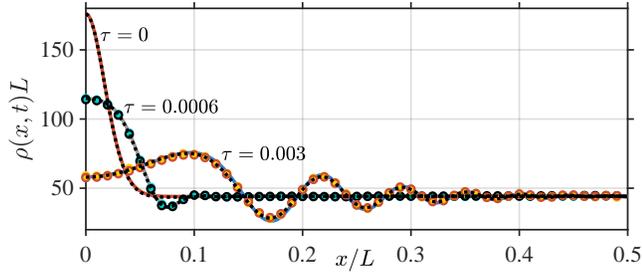

calculations, however, here we wanted to compare with other methods, specifically to reproduce background Friedel oscillations characteristic of finite-size systems or periodic potentials; with IBC, such oscillations would be absent for a single bump in an infinite (length-wise) background as the Friedel oscillation amplitude decays with increasing system size.

For obtaining the $t = 0$ initial state, we use the infinite density matrix renormalization group (iDMRG) algorithm, with $M = 1000$ sites per unit cell, keeping typically $10 - 200$ basis states with an energy variance $\sigma_E^2/J^2$ of $10^{-8} - 10^{-4}$ per site. For the time evolution, we use the infinite time-evolving block decimation (iTEBD) algorithm [S32], which is essentially equivalent to the Lie-Suzuki-Trotter form of adaptive time-dependent density matrix renormalization group (tDMRG) [S33, S34]. For most calculations presented here, we used the optimized 4th-order decomposition from Barthel and Zhang [S35], with time step 0.2 (measured in dimensionless units of time $Jt/\hbar$), and using a cutoff density matrix eigenvalue of $10^{-10}$, corresponding to a cutoff singular value of $10^{-5}$. We note that a cutoff value for the eigenvalues is more robust than fixing the truncation error (sum of discarded eigenvalues).

Errors in the TEBD algorithm arise from two sources, (1) the so-called *Trotter error*, arising from a finite time-step, and (2) truncation errors arising from the projection of the SVD to a finite bond-dimension. Thus the optimal time-step is an optimization between the truncation error and the Trotter error. To quantify the non-unitary errors arising from truncations, the most reliable method to determine the error is to perform the backwards evolution from the final time-evolved state, and verify that there is a high fidelity $F$ with the initial $t = 0$ state. For example, in the calculation of Fig. S7, the distance $d \equiv 1 - |F|$ with the $t = 0$ state after backwards evolution is $\sim 10^{-6}$ per site.

In Fig. S7 we show the iMPS results for the shock wave dynamics for the parameters of Fig. 1 (b) of the main text. As we see, all methods incorporating quantum fluctuation (positive-$P$, truncated Wigner, and iMPS), are in excellent agreement with each other, while also showing small departures from the mean-field GPE predictions, which are the largest near the local minima and maxima of the final-time ($\tau = 0.003$) density distribution (see main text for further details).

## VII. EXACT DIAGONALIZATION IN THE TONKS-GIRARDEAU REGIME

In the Tonks-Girardeau (TG) regime of infinitely strong interactions ($g \to \infty$), the Bose gas can be treated via a Bose-Fermi mapping [S36, S37] which reduces the dynamics of the many-body problem to the dynamics of single-particle states. In particular, the $N$-body wavefunction of the TG gas, $\Psi$, can be obtained via

$$\Psi(x_1, \ldots, x_N; t) = A(x_1, \ldots, x_N)\Psi^{(F)}(x_1, \ldots, x_N; t), \tag{S22}$$

where $\Psi^{(F)}$ represents the free fermion $N$-body wavefunction for the same dynamical scenario, and the unit antisymmetric function is given by $A(x_1, \ldots, x_N) = \prod_{1 \leq j < i \leq N} \text{sgn}(x_i - x_j)$ which ensures the correct symmetrisation of the bosonic wavefunction. Here, the sign function is given by,

$$\text{sgn}(x) = \begin{cases} -1 & \text{if } x < 0, \\ 0 & \text{if } x = 0, \\ 1 & \text{if } x > 0. \end{cases} \tag{S23}$$

The $N$-body fermionic wavefunction itself can be constructed using the Slater determinant

$$\Psi^{(F)}(x_1, \ldots, x_N; t) = \det_{i,j=1}^{N} [\phi_i(x_j, t)], \tag{S24}$$

where the single-particle wavefunctions $\phi_i(x, t)$ evolve according to the Schrödinger equation from their initial states $\phi_i(x, 0)$, which are eigenstates of the initial trapping potential $V(x, 0)$ with eigenenergies $E_i$ such that the total energy of the $N$-body wavefunction is $E_{\text{TG}} = \sum_{i=1}^{N} E_i$.

In general, the dynamical properties of the TG gas can be determined through the reduced one-body density matrix,

$$\rho(x, y; t) = \int dx_2 \ldots dx_N \\ \times \Psi(x, x_2, \ldots, x_N; t)\Psi^*(y, x_2, \ldots, x_N; t). \tag{S25}$$

In this Letter however, we are interested primarily with the evolution of the real-space density of the gas $\rho(x, t) = \rho(x, x; t)$. This can be found via

$$\begin{aligned} \rho(x, t) &= \rho(x, x; t) \\ &= \int dx_2 \ldots dx_N \, |\Psi(x, x_2, \ldots, x_N; t)|^2 \\ &= \int dx_2 \ldots dx_N \\ &\quad \times A^2(x, x_2, \ldots, x_N; t) \left|\Psi^{(F)}(x, x_2, \ldots, x_N; t)\right|^2 \\ &= \int dx_2 \ldots dx_N \left|\Psi^{(F)}(x, x_2, \ldots, x_N; t)\right|^2 \end{aligned} \tag{S26}$$

where $A^2(x, x_2, \ldots, x_N; t)$ can be dropped since it will simply multiply the fermionic wavefunction by $+1$. For the case where it would return 0, this corresponds to particles sharing

the same physical location and is hence already taken into account by construction of the fermionic many-body wavefunction which respects the Pauli exclusion principle.

Equation (S26) is exactly the real-space density $\rho^{(F)}(x,x;t)$ of the equivalent free fermion problem, and, upon substituting the Slater determinant (S24) into (S26) one can write the density of the bosonic TG gas in terms of the single-particle wavefunctions $\phi_i(x,t)$ as

$$\rho(x,t) = \rho^{(F)}(x,x;t) = \sum_{i=1}^{N} |\phi_i(x,t)|^2 , \qquad (S27)$$

where the cross terms which would appear from squaring the determinant vanish provided that the single-particle wavefunctions are orthonormal. Equation (S27) provides a simple way to compute the dynamics of the real-space density of the TG gas in terms of the single-particle wavefunctions of the noninteracting fermion problem.

After preparation of the initial many-body state, the dynamics of the single-particle wavefunctions themselves can be constructed using the expansion [S38]

$$\phi_i(x,t) = \sum_{n} c_n^{(i)} \psi_n(x) e^{-iE_n t/\hbar}, \qquad (S28)$$

where $\psi_n(x)$ ($n = 0, \pm 1, \pm 2, \ldots$) are the stationary eigenfunctions of the trapping potential under which the dynamics are occurring, with the $E_n$ being their respective eigenenergies. The $c_n^{(i)}$ are the expansion coefficients for the $i$th single-particle wavefunction $\phi_i(x,t)$. Since we have considered dynamics in a periodic box of length $L$ with uniform potential $V(x) = 0$, the eigenfunctions to be used here are plane waves,

$$\psi_n(x) = \frac{1}{\sqrt{L}} e^{ik_n x}, \qquad (S29)$$

with eigenenergies given by

$$E_n = \frac{\hbar^2 k_n^2}{2m} = \frac{2\pi^2 \hbar^2 n^2}{mL^2} . \qquad (S30)$$

Here $k_n = 2\pi n/L$ are the available wave numbers with $n = 0, \pm 1, \pm 2, \ldots$ being the allowed quantum numbers.

The only remaining unknown quantities from Eq. (S28) are the expansion coefficients $c_n^{(i)}$ which can be determined via the overlap integrals between the plane wave basis set $\psi_n(x)$ and the initial single-particle wavefunctions $\phi_i(x,0)$,

$$c_n^{(i)} = \int dx \, \psi_n^*(x) \phi_i(x,0). \qquad (S31)$$

The initial single-particle wavefunctions $\phi_i(x,0)$ that we use here are eigenstates of the initial trapping potential $V(x)$. Hence, in order to construct them, we would first like to determine the trap that will produce the desired initial density profile, given by Eq. (S7), as its ground state. Ideally, we would like this trap to be experimentally realistic and smooth. In the TG regime however, a smooth density profile (free of

Friedel oscillations), like that of Eq. (S7), can only be realised in a smooth trapping potential in the thermodynamic limit. For finite number of particles $N$, on the other hand, that same smooth potential will necessarily produce small-amplitude Friedel oscillations; modulations on top of the otherwise smooth thermodynamic limit density profile. For the sake of simplicity and definiteness, we go ahead and determine a realistically smooth trapping potential that corresponds to the desired initial density profile of Eq. (S7) in the thermodynamic limit, with the recognition that this trap will produce Friedel oscillations in our exact numerical examples (which are always for finite $N$). The task of finding such a trapping potential can be accomplished within the local density approximation [S4], using the local chemical potential $\mu(x)$, wherein the respective density profile in the thermodynamic limit is referred to as the Thomas-Fermi profile.

For a TG gas, the chemical potential is the same as that of an ideal 1D gas of spinless fermions, and for a uniform system at density $\rho$ it is thus given by $\mu = \hbar^2 \pi^2 \rho^2 / 2m$. For a trapped (nonuniform) system, one can introduce the local chemical potential $\mu(x) = \mu_0 - V(x)$, where $\mu_0$ is the global equilibrium chemical potential. In the local density approximation, $\mu(x)$ will simply be given by the same expression as in the uniform case except that $\rho$ is replaced by $\rho(x)$, i.e., $\mu(x) = \hbar^2 \pi^2 \rho(x)^2 / 2m$. Hence, in the thermodynamic limit of the TG gas, the trapping potential required to produce a desired density profile $\rho(x)$ is given by $V(x) = \mu_0 - \hbar^2 \pi^2 \rho(x)^2 / 2m$. Since $\mu_0$ is a constant and only provides a shift to the trap, it can be set to $\mu_0 = 0$ so long as it is ensured that the resulting density profile is correctly normalised to the total number of particles $N$. Using Eq. (S7) as an approximate guide for the initial density profile, this procedure results in

$$V(x) = -\frac{\hbar^2 \pi^2 \rho(x)^2}{2m} = -\frac{\hbar^2 \pi^2}{2m} N_{\text{bg}}^2 \left(1 + \beta e^{-x^2/2\sigma^2}\right)^4 . \qquad (S32)$$

For sufficiently large $N$, such a trapping potential will produce initial density profiles that are approximately the same as in the ideal Bose gas and GPE regimes of the main text (or equivalently, Eq. (S7) in dimensionless form). Hence, the relationship between $N_{\text{bg}}$ and $N$ from the normalization condition for $N \gg 1$ is approximately the same as in the main text, i.e., $N_{\text{bg}} = N / \left(1 + \frac{\sqrt{\pi}\beta\sigma}{L}\left[\beta \operatorname{erf}(\frac{L}{2\sigma}) + 2\sqrt{2} \operatorname{erf}(\frac{L}{2\sqrt{2}\sigma})\right]\right)$.

In dimensionless form, the trapping potential of Eq. (S32) can be written as

$$\bar{V}(\xi) = -\frac{\pi^2 \bar{\rho}(\xi)^2}{2} = -\frac{\pi^2}{2} N_{\text{bg}}^2 \left(1 + \beta e^{-\xi^2/2\bar{\sigma}^2}\right)^4 , \qquad (S33)$$

which can be compared to the trapping potential from Eq. (S9) used for the weakly-interacting regime in Sec. II. The trapping potential of Eq. (S33) used in the example of Fig. 2 (d) of the main text is shown here in Fig. S8 (a), whereas the one used in Fig. 2 (e) of the main text is shown here in Fig. S8 (b) and is given by Eq. (S9) with $\bar{g} = 0$.

Returning now to the construction of the initial single-particle eigenfunctions $\phi_i(x,0)$, these can be determined

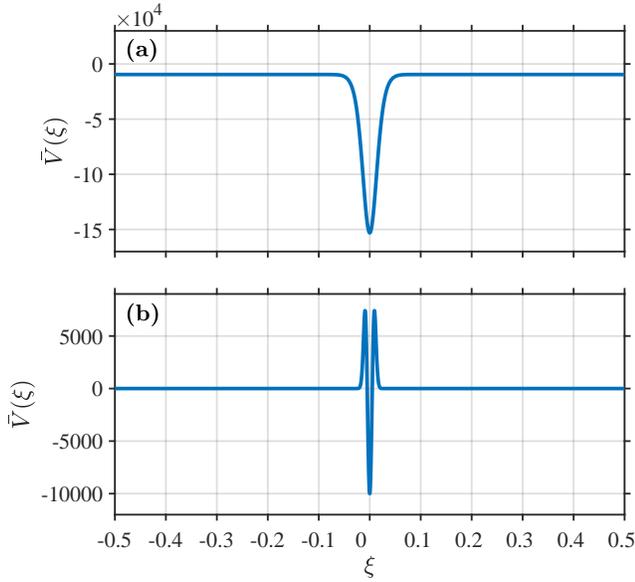

FIG. S8. Trapping potentials used, respectively, to prepare the initial ground state density profiles of Fig. 2 (d) and (e) of the main text.

through numerical diagonalization of the Hamiltonian

$$\hat{H} = -\frac{\hbar^2}{2m}\frac{\partial^2}{\partial x^2} + V(x), \tag{S34}$$

where $V(x)$ is given by Eq. (S32). After evolving these $\phi_i(x, 0)$ in time according to Eq. (S28), the time evolved density of the TG gas is then calculated using Eq. (S27). This is what was done to evaluate the density profiles shown in Figs. 2 (d) and (e) of the main text.

## VIII. COMPARISON OF THE RESULTS IN THE TONKS-GIRARDEAU REGIME WITH THE PREDICTIONS OF THE MODIFIED GPE AND THE RESULTING DISPERSIVE HYDRODYNAMICS

Using density functional arguments, Kolomeisky *et al.* [S39] have previously proposed a long-wavelength mean-field approach for the Tonks-Girardeau gas, which we refer to here as the modified Gross-Pitaevskii equation. Unlike the GPE which contains a quadratic interaction term, the modified GPE instead includes a quartic interaction term and is given by

$$i\hbar\frac{\partial\Phi(x,t)}{\partial t} = \left(-\frac{\hbar^2}{2m}\frac{\partial^2}{\partial x^2} + V(x) + \frac{\pi^2\hbar^2}{2m}|\Phi(x,t)|^4\right)\Phi(x,t). \tag{S35}$$

Here, $\Phi$ is similar to the order parameter in the mean-field GPE regime, though Kolomeisky *et al.* do not mention what this represents physically, yet they assume that the mean density of the TG gas is given by $\rho(x, t) = |\Phi(x, t)|^2$.

As with the GPE, a Madelung transformation can be performed on Eq. (S35) which results in a respective 'dispersive' hydrodynamic formulation. For ease of comparison with the

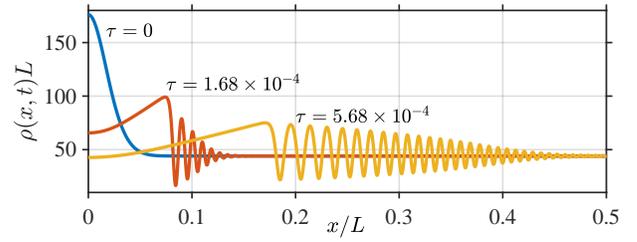

FIG. S9. Modified GPE prediction of dispersive shock waves in a Tonks-Girardeau gas. This mean-field prediction can be compared directly with the exact results of Fig. 2 (d) from the main text where $N = 50$, $\beta = 1$ and $\sigma/L = 0.02$. The initial density profile used here is the thermodynamic limit TG profile, *i.e.* the profile given by Eq. (S7), which results from using the trapping potential given in Eq. (S32).

GPE and superfluid hydrodynamics given in Sec. II, the dimensionless modified GPE can be rewritten as

$$i\frac{\partial\bar{\Phi}(\xi,\tau)}{\partial\tau} = \left(-\frac{1}{2}\frac{\partial^2}{\partial\xi^2} + \bar{V}(\xi) + \frac{\pi^2}{2}|\bar{\Phi}(\xi,\tau)|^4\right)\bar{\Phi}(\xi,\tau), \tag{S36}$$

with the resulting 'dispersive' hydrodynamics given by

$$\frac{\partial\bar{\rho}}{\partial\tau} = -\frac{\partial}{\partial\xi}(\bar{\rho}\bar{v}), \tag{S37}$$

$$\frac{\partial\bar{v}}{\partial\tau} = -\frac{\partial}{\partial\xi}\left(\frac{1}{2}\bar{v}^2 + \bar{V}(\xi) + \frac{\pi^2\bar{\rho}^2}{2} - \frac{1}{2}\frac{1}{\sqrt{\bar{\rho}}}\frac{\partial^2\sqrt{\bar{\rho}}}{\partial\xi^2}\right). \tag{S38}$$

As with standard superfluid hydrodynamics, given by Eq. (S5), the quantum pressure term appears again here as the last term in Eq. (S38). The usual pressure term though is represented by $-\partial_\xi(\pi^2\bar{\rho}^2/2)$; it corresponds to the equivalent normal pressure term $-\frac{1}{m\rho}\partial_x P$, present in the classical hydrodynamic equation for the velocity field. It originates from the thermodynamic equation of state for the pressure $P$ of an ideal 1D Fermi gas in a box potential (which is identical to that of the TG gas) given by $P = \hbar^2\pi^2\rho^3/3m$ [S10, S40, S41]. For comparison, the same normal pressure term in superfluid hydrodynamics, Eq. (S5), originates from the thermodynamic pressure $P = \frac{1}{2}g\rho^2$ of a weakly interacting 1D Bose gas in the mean-field regime.

Rather than postulating the modified GPE via density functional arguments, other authors [S42] have first postulated classical hydrodynamics for the TG gas and then added the quantum pressure term, allowing them to convert back and solve the modified GPE for computational simplicity. This is done with the recognition that the quantum pressure term must be small for the scenario under consideration, and that any small scale features arising in the dynamics cannot be trusted.

As pointed out by Girardeau and Wright [S43], the modified GPE overestimates interference phenomena in the TG gas, particularly in the scenario they considered—a split and recombined TG gas. We find that the same is true for the scenario we have considered; a density bump expanding into a non-zero background. The prediction of the modified GPE is




\begin{thebibliography}{89}
\bibitem [{Non()}]{Nonlinearity}%
  \BibitemOpen
  \href@noop {} {}\bibinfo {howpublished} {The term nonlinear interaction in
  the NLSE or GPE refer, respectively, to the Kerr (or $\chi^{(3)}$)
  interaction induced by the nonlinear optical material, or to the $s$-wave
  scattering interaction ($g$) of the atomic BEC. This should not be confused
  with the nonlinear terms (with respect to the density and velocity fields)
  that arise in the partial differential equations in the equivalent
  hydrodynamic formulation of the problem; such nonlinear terms in the
  hydrodynamic equations are present even in the absence of any Kerr or
  $s$-wave interactions.}\BibitemShut {Stop}%
\bibitem [{err()}]{errors}%
  \BibitemOpen
  \href@noop {} {}\bibinfo {howpublished} {In all stochastic simulations,
  performed using the XMDS software package of Ref.~\cite{xmds}, the averages
  are over 100,000 stochastic trajectories and the standard error on the mean
  density is smaller than the radius of the circles.}\BibitemShut {Stop}%
\bibitem [{Com()}]{Comment3}%
  \BibitemOpen
  \href@noop {} {}\bibinfo {howpublished} {For the parameters of
  Fig.~\ref{fig:single_particle_and_GPE}\,(b), $N_{\mathrm{bg}}\!=\!44.03$, and
  the dimensionless nonlinearity
  $\bar{g}\!=\!gmL/\hbar^2\!=\!\gamma_{\mathrm{bg}}N_{\mathrm{bg}}$ in the GPE
  is $\bar{g}\!=\!4.403$; similarly, for the parameters of
  Fig.~\ref{fig:single_particle_and_GPE}\,(c), $N_{\mathrm{bg}}\!=\!1761$ and
  $\bar{g}\!=\!17.61$.}\BibitemShut {Stop}%
\bibitem [{Sha()}]{Shape}%
  \BibitemOpen
  \href@noop {} {}\bibinfo {howpublished} {We note that the precise details of
  the shape of the initial density profile with a local bump do not affect our
  main results and conclusions.}\BibitemShut {Stop}%
\bibitem [{Tem()}]{Temperature}%
  \BibitemOpen
  \href@noop {} {}\bibinfo {howpublished} {The dimensionless temperature
  parameters $\overline{\mathcal{T}}$ and $\bar{T}\!=\!k_BT/E_0$ (with
  $E_0\!=\!\hbar^2/mL^2$) are related by
  $\overline{\mathcal{T}}=2\bar{T}/\bar{\rho}_{\mathrm{bg}}^2\!=\!2\bar{T}/N_{\mathrm{bg}}^2$.}\BibitemShut
  {Stop}%
\bibitem [{Int()}]{Intractable}%
  \BibitemOpen
  \href@noop {} {}\bibinfo {howpublished} {For the parameters of
  Fig.~\ref{fig:single_particle_and_GPE}\,(b), the iMPS results are in
  excellent agreement with the truncated Wigner and positive-$P$ results, and
  are omitted from the figure for clarity, whereas for the parameters of
  Fig.~\ref{fig:single_particle_and_GPE}\,(c), the iMPS method is
  computationally intractable due to the large number of particles in the
  system. The positive-$P$ approach for
  Fig.~\ref{fig:single_particle_and_GPE}\,(c) is similarly computationally
  intractable due to large sampling errors.}\BibitemShut {Stop}%
\end{thebibliography}
\end{document}